%% file: bdc_ieee.tex
\documentclass[journal,pdftex]{IEEEtran}

\usepackage{hyperref}
\usepackage{adjustbox}
\usepackage{natbib}
\ifCLASSINFOpdf
   \usepackage[pdftex]{graphicx}
\else
   \usepackage[dvips]{graphicx}
  \fi

\usepackage{amsmath}                            
\usepackage{amssymb}                            
\usepackage{amsthm}

\usepackage[letterpaper, margin = 0.75in, top = 0.75in]{geometry}

\usepackage{longtable}
\usepackage{array}
\newcolumntype{P}[1]{>{\centering\arraybackslash}p{#1}}
\usepackage{hhline}
\usepackage{algorithm}
\usepackage{algpseudocode}
\usepackage{xparse}
\usepackage{xcolor}

\makeatletter
\NewDocumentCommand{\LeftComment}{s m}{%
  \Statex \IfBooleanF{#1}{\hspace*{\ALG@thistlm}}#2}
\makeatother

\begin{document}

\title{Pass Evaluation in Women's Olympic Hockey} 

\author{\IEEEauthorblockN{Robyn Ritchie\IEEEauthorrefmark{1},
Alon Harell\IEEEauthorrefmark{2}, and Phil Shreeves\IEEEauthorrefmark{3}}
\\
\IEEEauthorblockA{Simon Fraser University, Burnaby, BC, Canada\\
Email: \IEEEauthorrefmark{1}robyn\_ritchie@sfu.ca,
\IEEEauthorrefmark{2}aharell@sfu.ca,
\IEEEauthorrefmark{3}phil\_shreeves@sfu.ca}}

\maketitle
\begin{abstract}
    Passing during power plays in hockey is a crucial component to move one's team closer to scoring a goal. With the use of women's ice hockey event and tracking data from the elimination round games during the 2022 Winter Olympics, we evaluate passing and assess players' risk-reward behaviours in these high intensity moments. We develop a model for probabilistic passing that accounts for the order of arrival to a desired location and potential interceptions along the way. This model is based on a player-specific motion model and a puck motion model that determines how far each player can reach in the time it takes the puck to get to a target. In addition, we model the rink control for each team and the scoring probability of the offensive team. These models are then combined into novel metrics for quantifying where a pass should be made such that it would result in a high scoring opportunity or result in a high chance of maintaining possession. Finally, we use various metrics to evaluate passes made throughout the available power plays and compare them to the optimal options at that time. This can be used to identify players' risk-reward tendencies and can be used by coaches when selecting which players are best suited for a power play given the circumstances of the game. 
\end{abstract}
\section{Introduction}
Key elements of a power play include maintaining puck possession and putting your team in an advantageous position to shoot on goal. Players that make the decision to pass need to do so quickly and strategically. They need to consider the speeds of all players and how hard to deliver the puck so that it will result in a completed pass and will not be intercepted along the way. In addition, while on a power play, a player would ideally be passing to a player that is in a good position to shoot and potentially score. As hockey is a fast paced game with many complex and moving parts, quantifying the various physical and psychological elements surrounding a pass is no small task. 

In this work, we investigate passing in women's Olympic power plays by determining the probability of interception, the ice control of each team, and the shot potential at each location on the ice. These complex metrics are then be combined to evaluate the quality of attempted passes, as well as compare to other possible passes. By contrasting the passes made and the available passes, we determine which players are making the best decisions on where to pass in order to put their team in a beneficial position and maintain possession. Next, by analyzing multiple passes per player, we evaluate players' risk-reward behaviours as well as their overall passing performance. 

\section{Data Preparation}
The data for this work has been provided by Stathletes and includes information on power play events and their associated tracking data for the 2022 Olympic women's elimination hockey games. Event data includes information on shots, takeaways, face-offs, passes, and much more along with the time, involved players' names, locations, and other relevant details. Tracking data was generated from broadcast video at 30 frames per second. 
For our work, we use data that has been cleaned by \href{https://github.com/the-bucketless/bdc}{\textcolor{blue}{The Bucketless}} and contains adjustments for coordinates, corrections for player and team mislabellings, added jersey numbers where missing values occur, and estimated track locations for players that jump in and out of tracked frames.

As our focus is on assessing pass quality in offensive plays, we restrict the data we use to include only direct passes in the offensive zone. For consistency across events, we convert all values so that they occur in the right half of the ice. To ensure we are evaluating situations fairly, we remove any passes where the tracking data consisted of only one team, or less than two players of either team. We were further able to increase the number of passing plays which met this criteria by incorporating the passer and receiver locations as defined in the event data if either player is out of the related camera shot. This gave us a total of 289 passes to evaluate.

\section{The Math Behind the Pass}
\begin{figure}[h]
  \centering
  \includegraphics[width=0.45\textwidth]{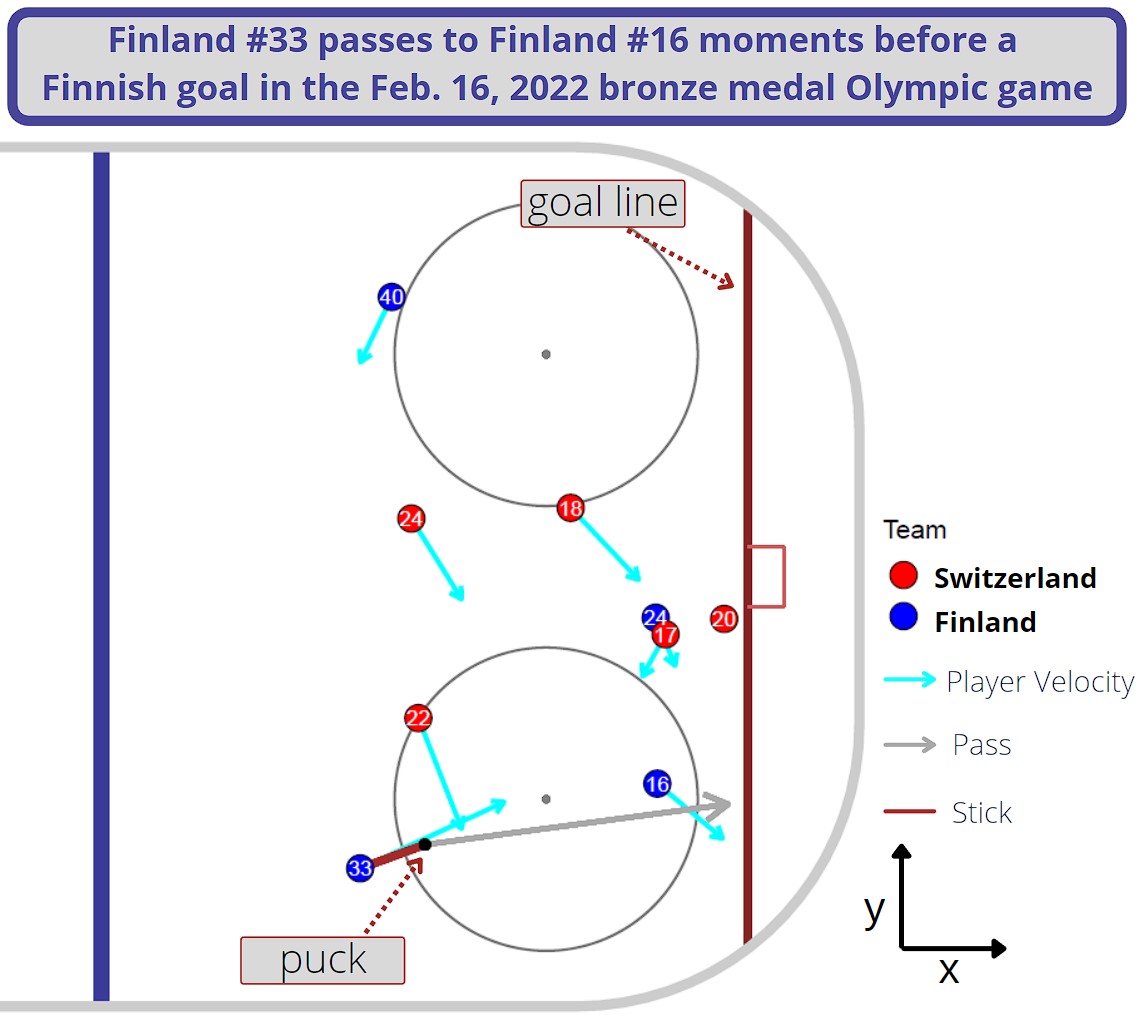}
  \caption{A pass (gray arrow) made by Finland against Switzerland in the third period of the Olympic bronze medal game.}
  \label{actual_pass}
\end{figure}

As we proceed through the math used to evaluate each pass, we use the example of the pass made on February 16, 2022 in the Olympic bronze medal game between Finland and Switzerland. In the third period, Finland's \#33 attempted a direct pass to Finland's \#16 as shown in Figure \ref{actual_pass}. The decision to make this pass was critical as it led to a Finnish goal shortly after. Throughout this paper, we show how this pass can be evaluated in various different ways to quantify the overall potential a pass to this location contains. 

\subsection{Motion Modelling}\label{subsec:motion}
In order to develop a proper passing evaluation model, we first model movement for both the players and the puck. \citet{fujimura2005geometric} provide a movement model that takes resistive force into consideration. This results in players having a maximum speed, along with a much more realistic movement in comparison to previous models~\citep{voronoi1908nouvelles,taki2000visualization}. We use this model to provide predictions for the positions of our players, $\mathbf{P}=(x,y)$ at a given time $t$, using~\eqref{eq:movement}. Here, a player starts at location $\mathbf{P_0}=(x_0,y_0)$, with initial speed $v = ({v_x,v_y})$, and begins accelerating in the direction $\phi$\footnote{$\phi$ is relative to the x-axis and within the interval of $(-\pi,\pi]$.} at $t=0$, with the assumption that she will be able to reach the maximum speed in any direction. The resulting position is
\begin{equation}
\scalebox{0.95}{$%
\begin{aligned}
    x &= x_0 + v_\text{max}\cdot\text{cos}(\phi)\cdot\Big(t-\frac{1-e^{-\gamma t}}{\gamma}\Big) + v_x\cdot\frac{1-e^{-\gamma t}}{\gamma}\text{, and}\\
    y &= y_0 + v_\text{max}\cdot\text{sin}(\phi)\cdot\Big(t-\frac{1-e^{-\gamma t}}{\gamma}\Big) + v_y\cdot\frac{1-e^{-\gamma t}}{\gamma}.
    \label{eq:movement}
\end{aligned}
$}
\end{equation}
Above, $\gamma=1.3$ sec$^{-1}$ is a parameter proportional to the resistive force, equal to the value used in~\citet{fujimura2005geometric}. Maximum speed\footnote{This is based on the maximum speed achieved by Coyne Schofield during the 2019 NHL All-Star fastest skater challenge.~\citep{cbcnews_2019}} is assumed to have a magnitude of $v_\text{max}=35.5$ ft/sec. The initial speed, $v$, is calculated for each player in each frame using $v=({v_x,v_y})=(\Delta x,\Delta y)\cdot f_\text{rate} $, where $\Delta x$ and $\Delta y$ are the differences in position between the current and previous frames, and $f_\text{rate} = 30$ Hz is the tracking data frame rate. In order to account for player reaction time, we assume that each player continues along their original path for $t_r=0.189$ seconds~\citep{lipps2011} before they begin moving according to~\eqref{eq:movement}. Note that we limit the movement of the goal keeper as we do not expect them venture away from the goal in an attempt to intercept the puck. We limit their movement to within eight feet of the net. 

We model the motion of the puck from the initial location, $(x_0,y_0)$, and speed, $(v_x,v_y)$, using a physical model accounting for both air resistance, using Stokes' drag \citep{stokes1850effect}, and friction from the ice,
\begin{equation}
\scalebox{0.95}{$%
\begin{aligned}
    x &= x_0 + \Big(v_x+\frac{\mu_g}{\kappa}\frac{v_x}{|v|}\Big)\cdot\Big(1-\frac{e^{-\kappa t}}{\kappa}\Big)-\frac{\mu_g}{\kappa}\frac{v_x}{|v|}t \text{, and}\\
    y &= y_0 + \Big(v_y+\frac{\mu_g}{\kappa}\frac{v_y}{|v|}\Big)\cdot\Big(1-\frac{e^{-\kappa t}}{\kappa}\Big)-\frac{\mu_g}{\kappa}\frac{v_y}{|v|}t,
    \label{eq:puck}
\end{aligned}
$}
\end{equation}
where $\mu_g=0.1$ us the coefficient of friction and $\kappa=k/m$, such that $m$ is the mass of the puck and $k$ is Stokes' drag coefficient~\citep{allain2020}.

\begin{figure}[h]
  \centering
  \includegraphics[width=0.45\textwidth]{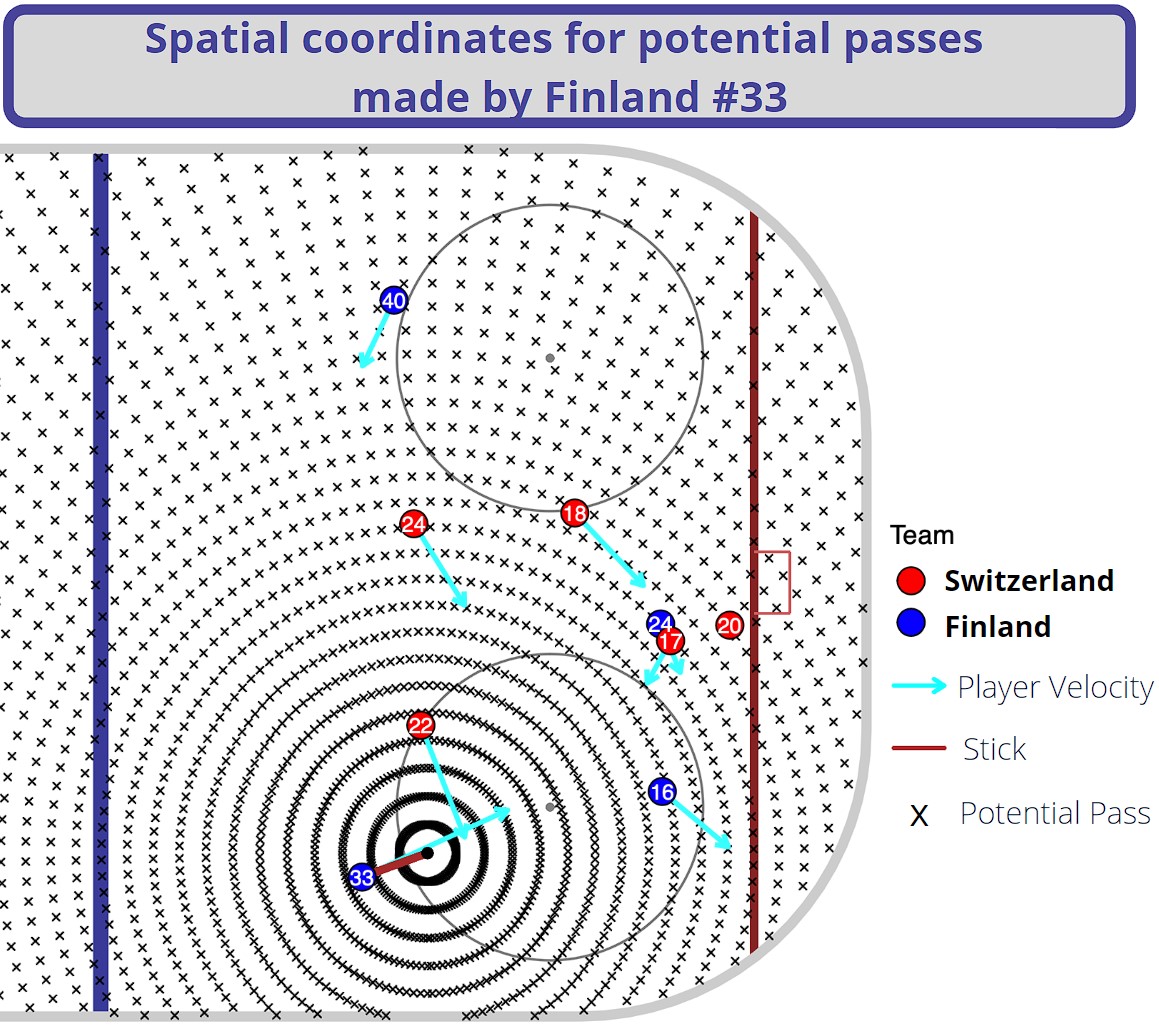} 
  \caption{All potential direct passes from the current puck location if it were sent in various different angles.}
  \label{potential_pass}
\end{figure}

Next, given the puck's current location and a pass speed, we evaluate where the puck will have been able to reach after some time, $t$, has passed. We do this at a regular time interval of 0.05 seconds, in all possible directions (at an angular resolution of 0.05 radians). This approach allows us to divide the ice into a grid of triplets of the form $Tr = (x,y,t)$, where $x,y$ are the spatial coordinates as shown in Figure \ref{potential_pass}. Additionally, any pass, in which the passer selects the angle and speed, can be described by a set of such triplets $\mathbf{Tr} = \{Tr_j:j\in1,2,...,K\}$ where $K$ is the number of time intervals until the puck hits the board.

\subsection{Rink Control}
Once models for both player and puck movement have been established, we quantify \textbf{rink control}~\eqref{eq:rink}; a measure of how much a team controls a certain region of the ice. This function, inspired by \cite{spearman2017physics} in soccer applications, stems from a physics model relating to the charge of a particle where $-1$ indicates a negative charge and $+1$ the opposite. We shift this scale from $[-1,1]$ to $[0,1]$, where $0$ indicates full control by the defence and 1 is indicative of full offensive control, resulting in
\begin{equation}\label{eq:rink}
\begin{small}
    \text{RC}(Tr_j) = \left[\frac{\sum_{i} l_i\cdot \tau_{ij}^{-\beta}}{\sum_i \tau_{ij}^{-\beta}} + 1\right]\cdot\frac{1}{2}.
\end{small}
\end{equation}
Here $l_i$ is a team label (+1 for offence and -1 for defence) corresponding to player $i$; and $\tau_{ij}$ measures the delay between the time it takes for player $i$ to reach the point $(x_j,y_j)$ and the time the puck arrives there, $t_j$. To calculate $\tau_{ij}$, we first calculate the time $t_{ij}$ it will take player $i$ to arrive at the desired point using the motion modelling described above, and then subtract the puck's arrival time $t_j$. We use $\beta$ as a parameter to measure the importance of being the first to reach the puck, where $\beta=0$ yields a RC equal to the proportion of offensive team players on the ice and $\beta=\infty$ yields a RC of 1 if the closest player belongs to the offensive team and 0 otherwise. The selected value of $\beta=2.5$ is borrowed from the original soccer application \citep{spearman2016pitch}.

\begin{figure}[h]
  \centering
  \includegraphics[width=0.45\textwidth]{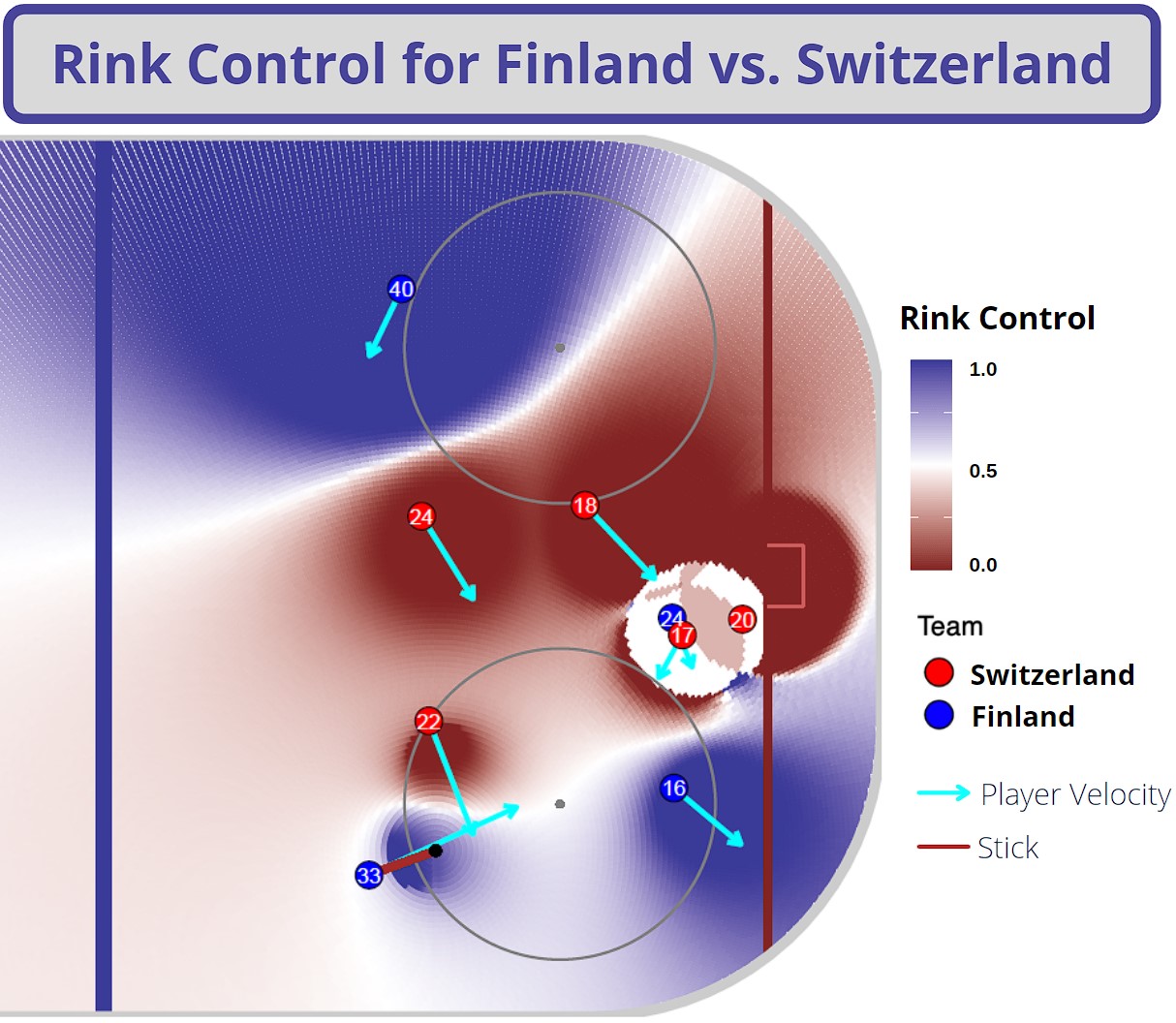} 
  \caption{Rink control for the current frame of the play. Dark red indicates a zone with high Swiss control and dark blue indicates a high Finnish control.}
  \label{rink_control}
\end{figure}

\subsection{Probabilistic Passing}
As explained above, the trajectory of a pass can be described by a set of triplets $Tr_j =\{(x_j,y_j,t_j) : j\in1,2,...,K\}$. At any triplet $Tr_j$ on the ice, we use the player and puck motion models to determine the probability that each player is able to pick-up the puck while accounting for all other players. These probabilities allow us to evaluate the current situation in two distinct ways. First, we assess the pick-up probability for each player assuming the puck reached a point of interest. Then, we follow along the trajectory of the pass and take into account the possibility of the pass being intercepted along the way. 

We begin by taking in the current location and velocities of each player and considering what would happen if they all converged on a triplet $Tr_j =(x_j,y_j,t_j)$. Using the player motion model, we determine how close each player $i$ could get to the point of interest by the time $t_j$ (when the puck arrives), this final player specific distance is denoted by $d_{ij}$. We associate a pick-up probability density around the puck using a normal distribution with a standard deviation defined by players' reach and assign probability to each player assuming they ``cover" an area of one reach in either direction over a time of $0.1$ seconds (and normalize accordingly to our temporal resolution). Finally, we add an \textit{exponentially decaying penalty} for intercept reaction time, $t_I$, which we assume to be $t_r$ for the offence and $t_r+0.1$ for the defence, to account for the defence reacting where the offence might be able to anticipate. Therefore player $i$, with distance to point $Tr_j$ identified by $d_{ij}$, has a base pick-up probability of
\begin{equation}\label{eq:pick-up potential}
\begin{small}
P^{(\text{base})}_{d_{ij}} =\left[\Phi\left(\frac{d_{ij}+s}{s}\right) - \Phi\left(\frac{d_{ij}-s}{s}\right)\right]e^{-t/t_I},
\end{small}
\end{equation}
where $\Phi(\cdot)$ is the CDF of the standard normal distribution and $s = 6.5$ ft is the player reach accounting for average height of women, stick length, arm length, and skate added height.

After calculating the distances, $d_{ij}$, and base pick-up probabilities, $P^{(\text{base})}_{d_{ij}}$, for each player present in the given frame, we update the pick-up probabilities to account for the order at which the players would arrive at the desired point. The ordered pick-up probabilities are calculated by ranking the players by their distances, $d_{ij}$. Let $R_{1j}$ indicate the event that the first player to reach the point of interest controls the puck, $R_{2j}$ indicate the event that the second player to arrive does so, and so on. These probabilities will now account for order of arrival in the sense that each player to arrive at the point of interest earlier must have failed to pick-up the puck in order for the current player to have a chance at it,
\begin{equation}\label{eq:ordered pick-up potential}
\begin{small}
    P_{R_{ij}|R_j}=
    \begin{cases}
    P^{(\text{base})}_{d_{ij}}\qquad &i=1\text{, or}\\
    P^{(\text{base})}_{d_{ij}}(1-\sum_{l=1}^{i-1} P_{R_{lj}|R_j})\qquad& i\neq 1.
    \end{cases}
\end{small}
\end{equation}
Note that these probabilities are conditioned upon the puck arriving at the desired point along the trajectory, which we denote by $R_j$. 

\begin{figure}[H]
  \centering
  \includegraphics[width=0.45\textwidth]{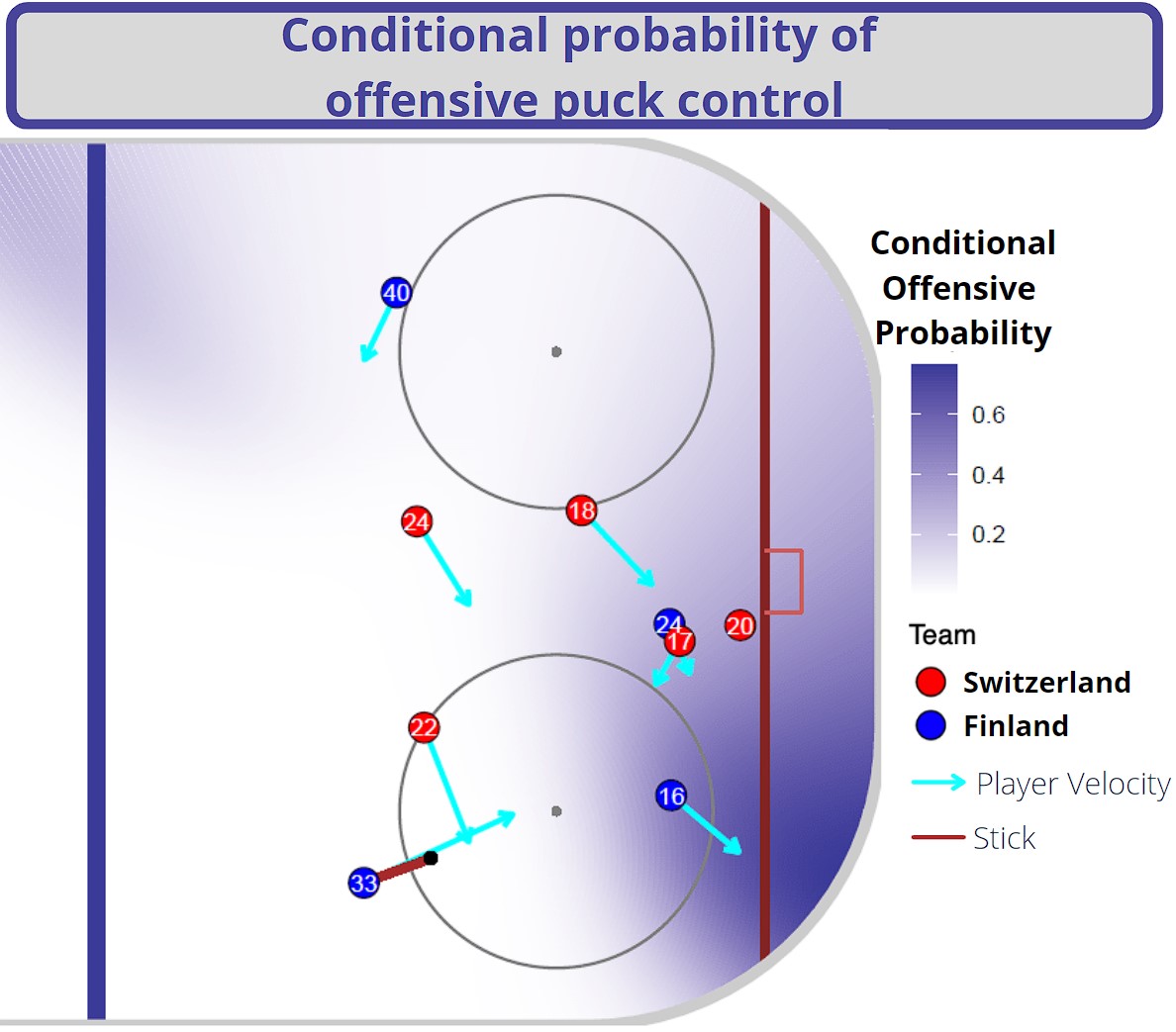} 
  \caption{Offensive probability of puck control conditional on the puck arriving to the particular point on the ice. This accounts for the order of arrival of the players to the puck at the location of interest.}
  \label{off_prob}
\end{figure}

At each triplet $Tr_j$, we are able to sum over the different players on each team (omitting the passer) to obtain team-wise probabilities. Summing
$P_{R_{ij}|R_j}$~\eqref{eq:ordered pick-up potential} over all players $i$ in the offence or defence gives the conditional probability of offensive control or defensive interception at any triplet on the ice (conditioned on the puck arriving at $Tr_j)$, which we denote $P_{\text{Off}|R_j}(Tr_j)$, shown in Figure \ref{off_prob}, and $P_{\text{Def}|R_j}(Tr_j)$, respectively.

Next, we follow the same steps as described previously, and account for whether or not the puck will reach the point of interest. For example, if there is a high probability of interception close to where the puck originates, then the chance of the puck making it to a further point along the same trajectory should be reduced. We accomplish this by following a similar rationale of the ordered pick-up probabilities~\eqref{eq:ordered pick-up potential}. Before a pass can arrive to point $j$, each player first gets a chance to intercept it at all the points $1,2,...,j-1$. Thus the final pick-up probability for player $i$ and location $Tr_j$ is defined by
\begin{equation}\label{eq:final trajectory pick-up probability}
\begin{small}
    P_{R_{ij}}=
    \begin{cases}
    P_{R_{ij}|R_j}\qquad &j=1\text{, or}\\
    P_{R_{ij}|R_j}(1-\sum_{l=1}^{j-1}P_{R_{il}})\qquad &j\neq 1.
    \end{cases}
\end{small}
\end{equation}

\begin{figure}[h]
  \centering
  \includegraphics[width=0.45\textwidth]{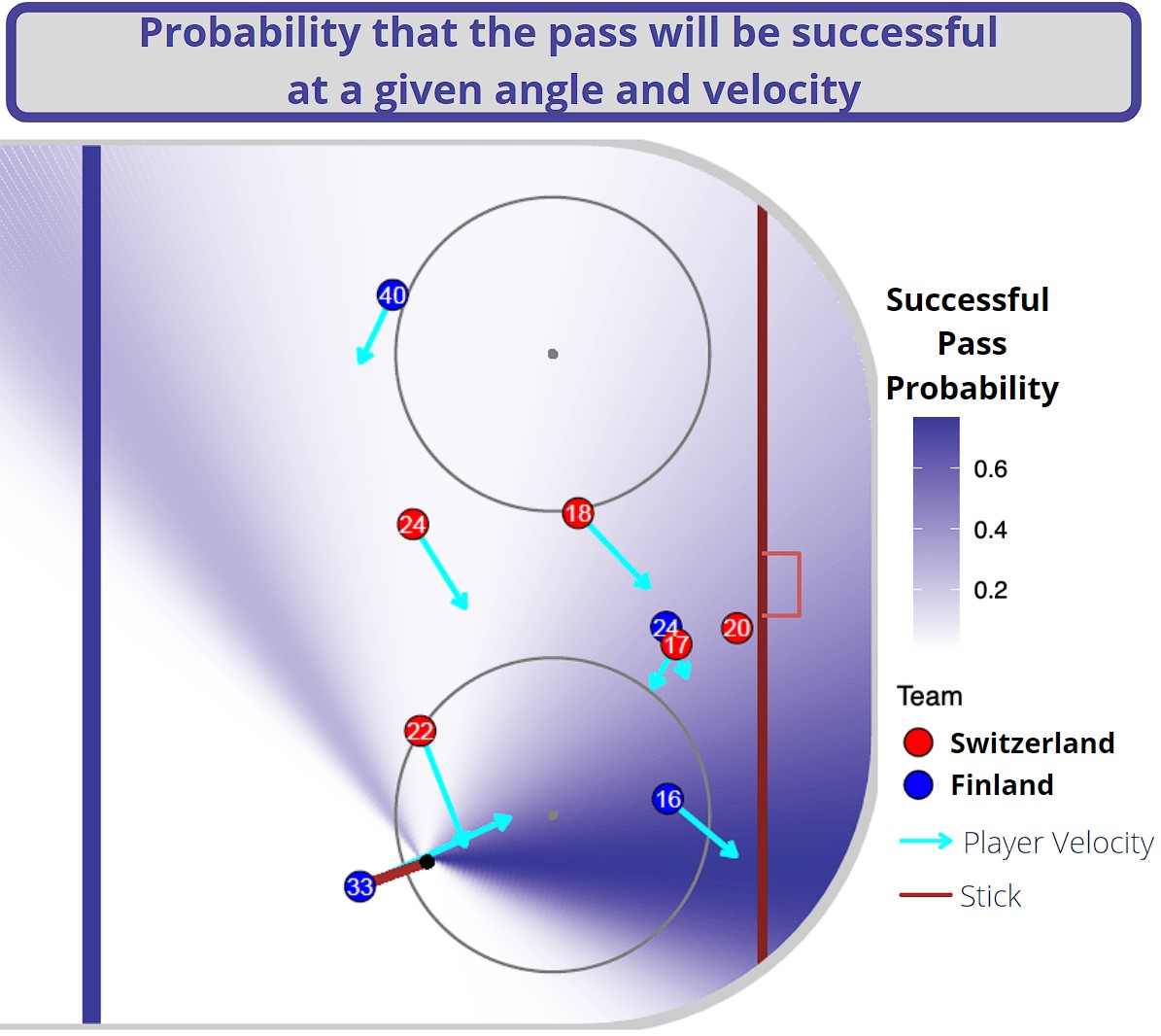} 
  \caption{Successful pass probability gives the player an idea of which angles will most likely result in a successful recovery by their teammate anywhere along a chosen angle.}
  \label{successful_pass}
\end{figure}

Summing over $P_{R_{ij}}$~\eqref{eq:final trajectory pick-up probability} removes the conditioning on the puck arriving.  This gives us the probability of offensive control (and defensive intercept) at \textit{each specific triplet} along the trajectory of a pass, which we denote $P_\text{Off}(Tr_j)$ and $P_\text{Def}(Tr_j)$. Finally, summing $P_\text{Off}(Tr_j)$ over all triplets $j$ that make up a pass in the direction $\alpha$ and speed $v_\text{pass}$, we get our final estimate of the probability\footnote{Note that because we look at a pass only over a limited number of time steps (until it reaches the boards), there may be some probability of the pass not being either controlled nor intercepted. We assume this is not a desired outcome, and only consider $P(\text{Success}|v_\text{pass},\alpha)$ as the probability of a successful pass.} of a successful pass, $P(\text{Success}|v_\text{pass},\alpha)$, as seen in Figure \ref{successful_pass}. 

\subsection{Scoring Probability}
\normalsize
To model the probability of scoring at a given point on the ice, a function that considers both the distance and angle with respect to the goal is required. We model the \textbf{scoring probability} with respect to distance using a function commonly referred to as the \textit{squared exponential function} in Gaussian process modelling~\citep{williams2006gaussian}. In those applications, it is used to model the correlation between points with respect to their inputs. This can be seen as a similar case, as we want to measure the correlation between a point of interest on the ice and the centre of the net. The further one deviates from the exact centre of the goal, the more difficult it is to score. Then prior to considering the angle of the shot, we express the scoring probability as
\begin{equation}
        \begin{aligned}
            \label{eq:sqexp}
            \widehat{\text{ScPr}}(x,y)=  \exp{\left[-\left(\frac{(x-189)^2}{\ell_x} + \frac{(y-42.5)^2}{\ell_y}\right)\right]},
        \end{aligned}
    \end{equation}
\normalsize
where $\ell_x$ and $\ell_y$ are parameters describing the decay in scoring probability with respect to the distance from the goal. Using prior knowledge regarding scoring probabilities in hockey (i.e. it is much more difficult to score near centre ice or on the boards rather than right in front of the goal), we set these values to be $\ell_x=2000$ and $\ell_y=500$, and assume that this applies equally to all players.

\begin{figure}[h]
  \centering
  \includegraphics[width=0.45\textwidth]{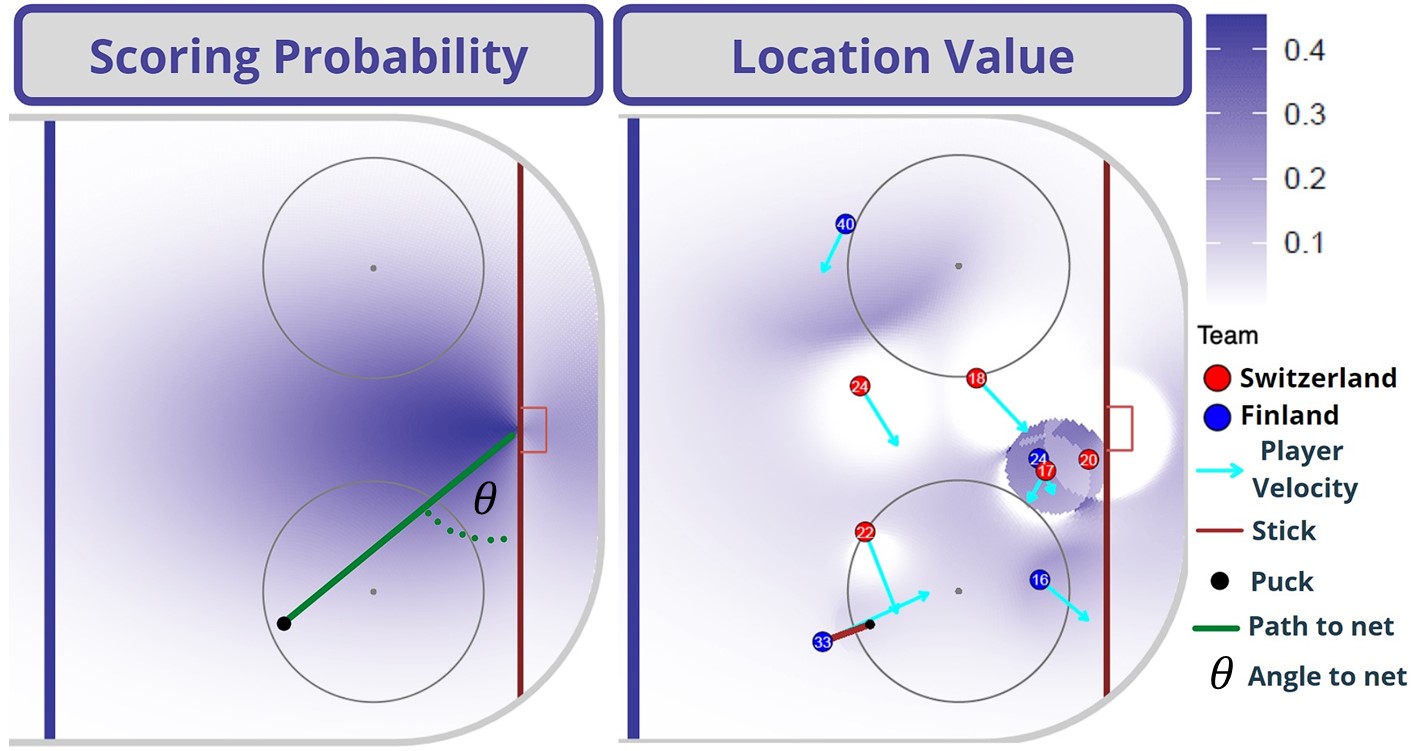} 
  \caption{Scoring probability (left) on a given location relative to their distance and angle from the centre of the net. Location value (right) is the result of multiplying the scoring probability with rink control.}
  \label{score_prob_loc_value}
\end{figure}

Next we consider the angle with respect to the goal, as it is more difficult to score when one is on the goal line, directly to the left or right of the net, than it is when they are directly in front of the goal. This is handled by multiplying $\widehat{\text{ScPr}}$ by $(\text{sin}(\theta)+1)/4$, where $\theta$ is the angle between the goal line and the line from the point of interest to the centre of the goal as shown in the left image of Figure \ref{score_prob_loc_value}. It should be noted that we scale the adjustment to angle in a similar fashion to rink control; this is to ensure there are no negative scoring probabilities when one is behind the net. Areas behind the goal line have an additional penalty, having their $\widehat{\text{ScPr}}$ value further multiplied by $1/2$. Note that the scoring probability behind the net is not zero, as we also consider the potential of a scoring play arising from this position. The resulting final scoring probability model (accounting for both angle and distance), is
\begin{equation}
        \text{ScPr}(x,y) = \widehat{\text{ScPr}}(x,y)\cdot
        \begin{cases}
            \frac{\text{sin}(\theta)+1}{4}  & \, x<189\text{, or}\\
            \frac{\text{sin}(\theta)+1}{8}  & \, x\geq189.
        \end{cases}\label{eq:scpr}
\end{equation}

The formulation above is of course somewhat naive in that it does not yet take into account the effects of the defence on the scoring probabilities. In order to account for the defensive pressure at any potential point we multiply the scoring probability~\eqref{eq:scpr} by the rink control~\eqref{eq:rink} to obtain what we call \textbf{location value},
\begin{equation}\label{eq:location value}
    \text{LV}(Tr_j) = \text{ScPr}(Tr_j(x), Tr_j(y))\cdot \text{RC}(Tr_j).
\end{equation}
The result can be seen in the right image of Figure~\ref{score_prob_loc_value}.

\section{The Pass Evaluation Model}
Having constructed the necessary tools in the previous section, we now begin to join them together in order to estimate the value a pass might have in a real game setting. We do this by observing the location values along the trajectory of a pass, and accounting for various aspects of our probabilistic passing model. This leads to a few different metrics, which together allow us to analyze the passing opportunities in a given game situation. Next, we look at the passes made by each player to analyze their passing and decision making tendencies.

\subsection{Passing Metrics}
\begin{figure}[h]
  \centering
  \includegraphics[width=0.45\textwidth]{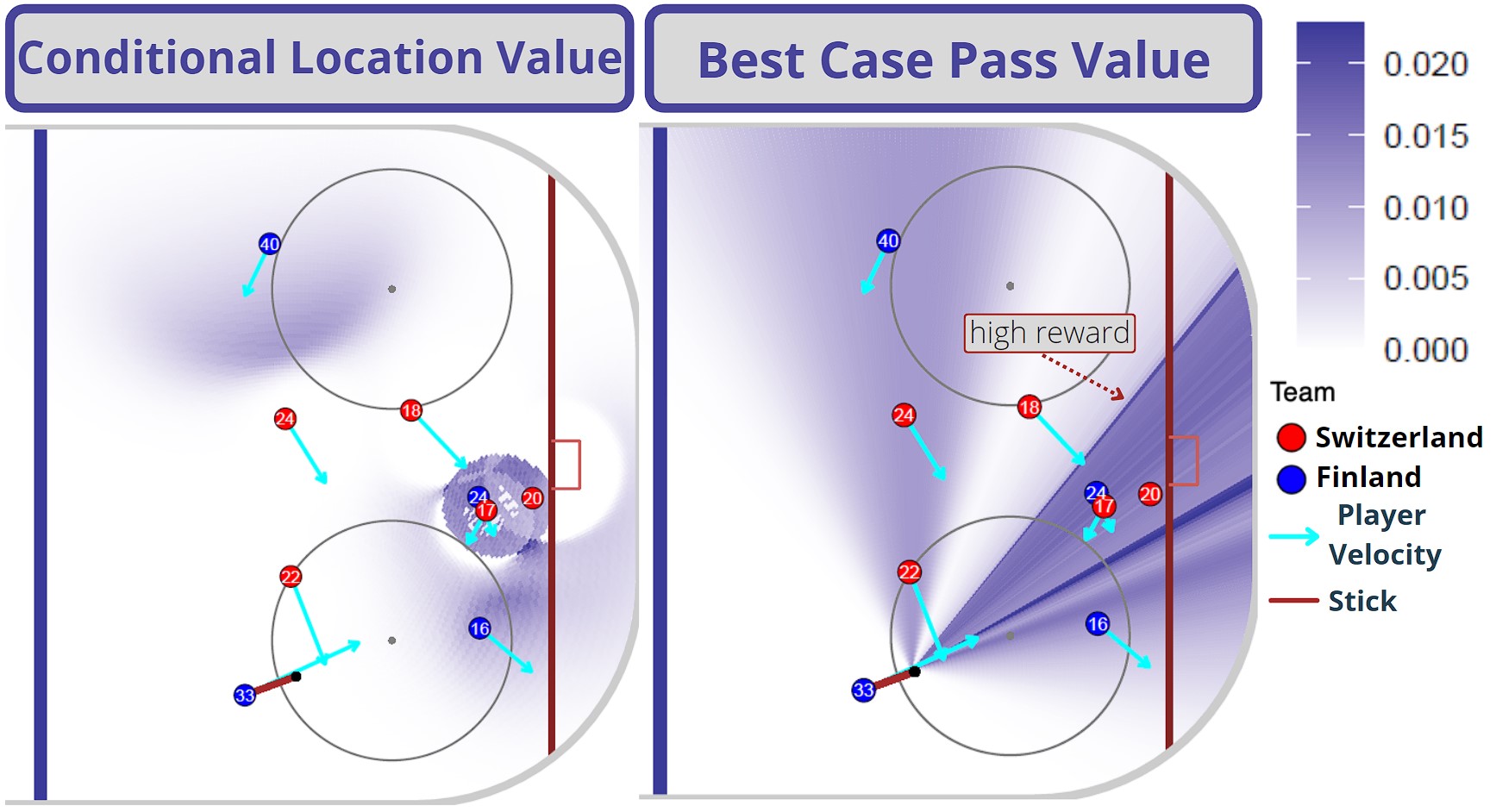} 
  \caption{Conditional location value (left) describes the locations on the ice that are under offensive control and have a high probability of reaching the desired point. The highest conditional location value along any angle and speed gives the best case pass value (right). This accounts for control of the ice and shot potential but not interceptions.}
  \label{fig:CLV_Best}
\end{figure}

Location value, shown in Figure~\ref{score_prob_loc_value} (right side), can be thought of as a metric for the offensive benefit of having the puck in a certain triplet ($Tr_j$, or location and time), when accounting for the rink control of that triplet. To incorporate this into a pass evaluation model, we first account for the probability that an offensive player will be able to control the puck there (and thus have a chance at scoring), assuming that the passer will be able to get the puck to this location. We call this \textbf{conditional location value},
\begin{equation}\label{eq:CLV}
    \text{CLV}_{R_j}(Tr_j) = \text{LV}(Tr_j)\cdot P_{\text{Off}|R_j}(Tr_j).
\end{equation}
We do this for all the points along the trajectory of any pass we wish to evaluate; thus explaining the use of $Tr_j$ to denote the location-time triplet (see left image of Figure~\ref{fig:CLV_Best}). 

\begin{figure}[h]
  \centering
  \includegraphics[width=0.45\textwidth]{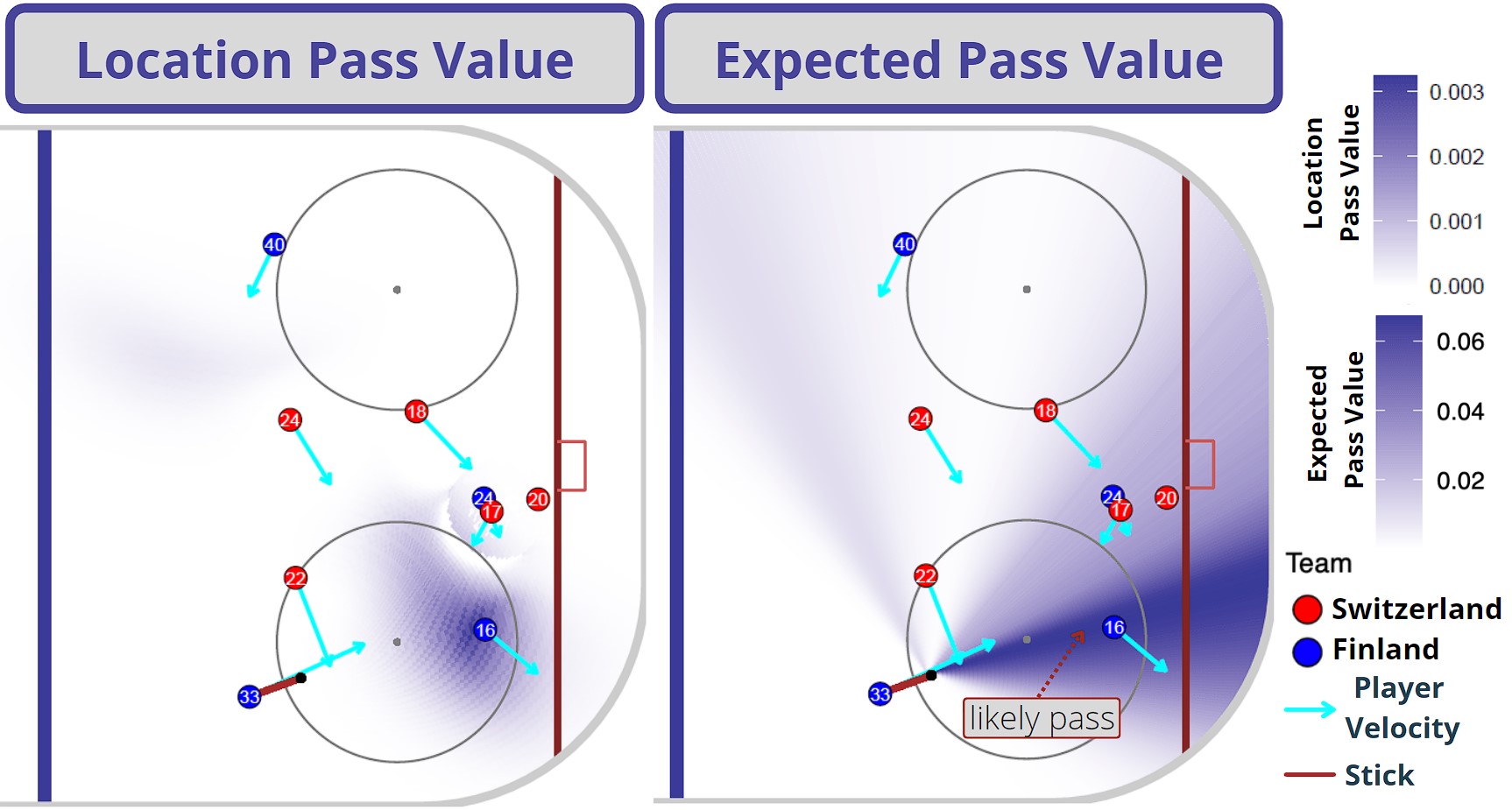} 
  \caption{Location pass value (left) is a metric which accounts for interception, offensive pass control, rink control and scoring probability. Summing over the location pass values along any angle describes the overall quality of a pass in such direction at the speed $v_\text{pass}$ and gives the expected pass value (right).}
  \label{fig:LPV_Expected}
\end{figure}

Next, we want to also account for the probability of the puck making it along the path to the desired triplet, which we do by replacing $P_{\text{Off}|R_j}(Tr_j)$ with $P_\text{Off}(Tr_j)$. We call this metric \textbf{location pass value} and it is given by
\begin{equation}\label{eq:LPV}
    \text{LPV}(Tr_j) = \text{LV}(Tr_j)\cdot P_\text{Off}(Tr_j).
\end{equation}
This metric represents our most comprehensive estimate of the value of each point along a pass trajectory (Figure~\ref{fig:LPV_Expected}, left).

The metrics above are extremely useful in analyzing the different components that make up a quality pass. However, in order to make them directly valuable to a player or a coach, we relate them to the aspects that are under the direct control of the player when they pass the puck - the speed of the pass and the direction in which they aim. We do this by grouping together the triplet-specific metrics above across all triplets of a pass. Note that for both metrics above (and the following metrics), the values of $Tr_j$ depend on the angle $\alpha$ and speed $v_\text{pass}$ of the analyzed pass, as explained in section~\ref{subsec:motion} .

Due to the fact that the conditional location value $\text{CLV}_{R_j}(Tr_j)$~\eqref{eq:CLV} measures the value of each point assuming the passer can get the puck there, we consider it a measure of the best case scenario of a pass. Thus, the metric measuring the best case scenario of an entire pass, which we call \textbf{best case pass value}, is simply the maximum of all $\text{CLV}_{R_j}(Tr_j)$~\eqref{eq:CLV} values over a potential pass' trajectory (see right image of Figure~\ref{fig:CLV_Best}). 

The location pass value $\text{LPV}(Tr_j)$~\eqref{eq:LPV} already accounts for the probability of the puck arriving at the triplet $Tr_j$ and thus in order to summarise it for a specific pass, we simply sum over all triplets along a given angle. We refer to this sum as the \textbf{expected pass value} and it represents our overall estimate of the quality of any single pass (see right image of Figure~\ref{fig:LPV_Expected}).

The best case pass and expected pass values, along with the successful pass probability, are extremely useful metrics. At any frame of the play, they allow for review of all potential passes where we can quantify where the optimal pass is depending on the situation at hand. If you are looking for a high reward, and want to increase your scoring opportunity, even at some risk of losing the puck, you might prioritize the pass with the highest best case pass value. If maintaining possession is key, you may instead prioritize the successful pass probability. Of course, if you are simply after the optimal available pass overall, using the expected pass value is ideal. Furthermore, by observing these values for the passes that were executed in practice, we can evaluate each player's tendencies as well as overall performance.

\subsection{Passing Analysis}\label{subsec:analysis}
One important characteristic of decision making is the trade-off between risk and reward. Some players favour the risky passes that may lead to a goal, over the safe passes that will maintain puck control, whereas other players may be more conservative. In order to analyze such tendencies, we look at the entire available body of work of each player and calculate their average best case pass value as well as their average successful pass probability. By plotting these two values against one another, we can immediately obtain a visual interpretation of their passing tendencies.

\begin{figure}[h]
  \centering
  \includegraphics[width=0.45\textwidth]{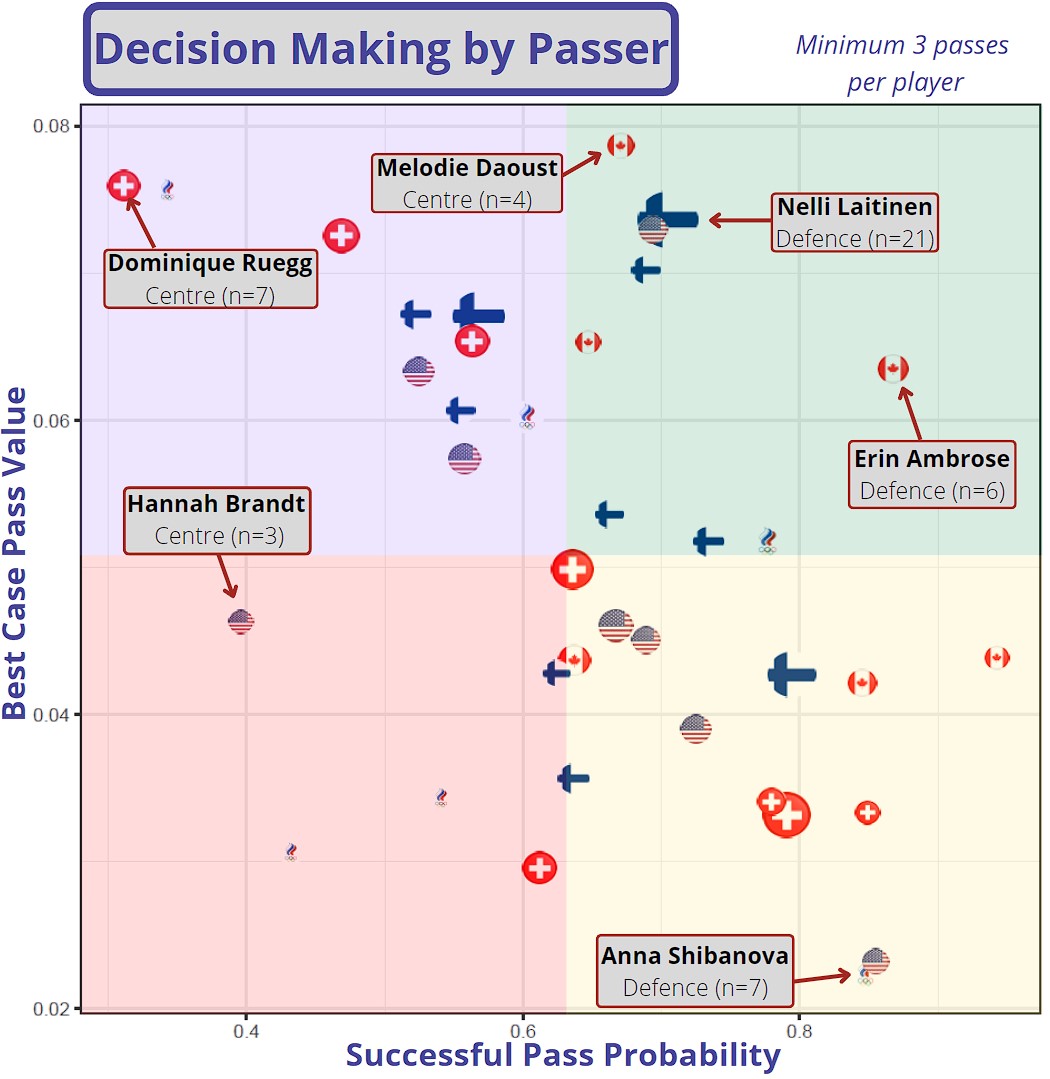} 
  \caption{The four quadrants of the decision making plot represent four distinct characteristics of passers: the green quadrant is best (low-risk high-reward); the yellow quadrant is conservative (low-risk low-reward); the purple quadrant is aggressive (high-risk high-reward); and the red quadrant is worst (high-risk low-reward).}
  \label{fig:decision}
\end{figure}

In the decision making plot, shown in Figure~\ref{fig:decision}, the top right represents the optimal combination of relatively safe passing with high best case values. The bottom right represents more conservative passing because, although they achieve a high probability of completion, they don't offer a very high best case value. The top left represents high-risk, high-reward passing, where the probability of completion may be quite low, but the value (if completed) can be quite high. Lastly we have the lower left, which represents poor passing both in terms of likelihood of completion as well as potential value. It is important to note that the sample size for this analysis is small, with some players only attempting as few as 3 passes, and our results would be more meaningful with larger quantities of data.

In order to evaluate overall passing performance, we quantify a pass' overall merit when compared with the player's alternatives. To do this, we first define the best passing outcome for a play as the maximum expected pass value over all possible passing angles and 4 possible pass speeds\footnote{These speeds were selected based on summary statistics of the attempted passes throughout the data. They are chosen to represent a soft, medium, and hard pass.}: 45, 65, 85 ft/sec, and the actual speed at which the pass was made\footnote{The actual passing speed is inferred from the event data, based on our puck motion model.}. Because a player may also choose to hold on to the puck or to shoot, we include the option of not passing in the maximum (to which we assign the location value at the passer's current position) and define it \textbf{best outcome}. We designate the \textbf{relative outcome} of a play to be the ratio between the best outcome and the expected pass value achieved in the actual play.\footnote{To account for inaccuracies in the tracking data, we assign the actual play an expected pass value which is the maximum of the values at a small range of angles around the actual pass direction, calculated using the actual pass speed.}

\begin{figure}[h]
  \centering
  \includegraphics[width=0.45\textwidth]{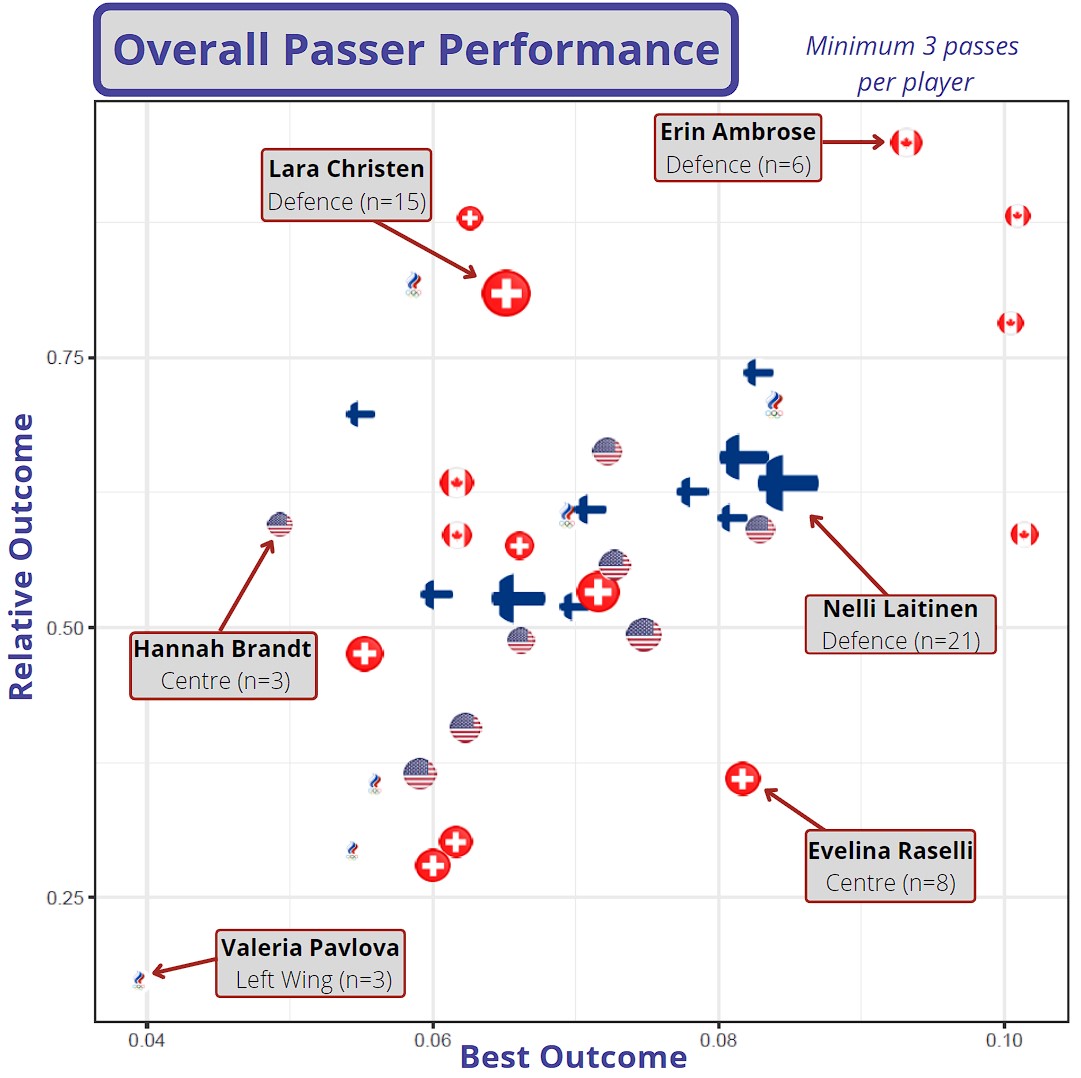} 
  \caption{Best outcome here measures the best alternative available in each passing situation. Relative outcome measures how well a player capitalizes on the potential value of the passes available to them.}
  \label{fig:overall_pass}
\end{figure}

Relative outcome allows us to analyze how well a player selects and executes the correct pass in a given scenario, whereas best outcome lets us know how good their available options were. An average relative outcome of 1 means that the player always chose the best pass available, and executed it well, whereas a low value indicates a sub-optimal pass. A high average best outcome means the player often put themselves in situations to make valuable passes, or perhaps, that their teammates' movement offered them good options for passes. The player and the coaching staff can then review individual plays, using our metrics, to determine the source of the difference between the best outcome and the pass made by the player. 

Figure~\ref{fig:overall_pass} shows the average relative outcome compared with the average best outcome for all players with at least 3 passes within our data. Here a player on the top right (such as Erin Ambrose of Canada) is one that had highly valuable passing opportunities and was able to capitalize on them; a player on the top left (such as Lara Christen of Switzerland) may not have had very valuable passes available to them, but made the most of their tough situation; a player in the lower half of the plot failed to maximise the value of their opportunities, if the best outcome was high this might represent a loss of a great chance.

\section{Summary}
In this work we have presented a robust framework for analyzing direct offensive passes in women's Olympic hockey. We began by defining our physical and probabilistic modelling, which then allowed us to build a multitude of performance metrics for each pass. We demonstrated how one might use the variety of our metrics to analyze a player's decision making and overall pass quality for any single play.  Coaching staff can use this in conjunction with film study to determine what the correct decision was, given the state of the game, and communicate this with the players using the illustrative graphics created. Because power plays are specifically unbalanced moments within a game, such an in-depth situational analysis of passing is particularly beneficial to both coaches and players. 

Following this, we created summary metrics, based on the dataset of passes available, to analyze each player's tendencies and overall quality of passes. We limited our summary metrics to players but one can easily generalize this work to team-level summaries. In addition to evaluating the best overall passers, we are able to identify risk-inclined players as opposed to conservative ones. This knowledge of players' passing characteristics can be used in a variety of beneficial ways. On a tactical level, it allows coaching staff to better prepare their team to an upcoming opponent's passing style; within a game, it gives a coach another important tool in deciding which players are most needed on the ice based on the current score, time and opponent; and for a more long-term view, it can be used by head office executives and scouting personnel to aid in decision making with respect to roster building.


To improve our accuracy, we can look to optimize our parameters using more specific details such as player heights, skating speeds, etc. It should also be noted that small sample size has a heavy impact on results and the incorporation of more data may influence player evaluation results. Additional tracking and event data would also allow for movement modelling, probabilistic passing, and scoring probabilities to account for individual player tendencies. For example, \cite{brefeld2019} provide probabilistic player movement models, which take player traits and movements into consideration and would be the next logical step for player movement development. Similarly, scoring probability maps could also be created for each player and game scenario, based on data from more shots and goals. Puck movement modelling can also be improved by accounting for it either bouncing off of or sliding along the boards. One particularly exciting use of our pass metrics, would be to include them as predictors in Markov chain modelling or reinforcement learning models. Such a study would allow the analysis of passes to be part of a larger whole of player actions, which together aim to maximise the long term scoring probability over multiple plays.

\newpage
\input{references.tex}

\end{document}

%% file: references.tex
\bibliography{bdc_ieee.bib}

%% file: bdc_ieee.bbl
\begin{thebibliography}{11}
\providecommand{\natexlab}[1]{#1}
\providecommand{\url}[1]{\texttt{#1}}
\expandafter\ifx\csname urlstyle\endcsname\relax
  \providecommand{\doi}[1]{doi: #1}\else
  \providecommand{\doi}{doi: \begingroup \urlstyle{rm}\Url}\fi

\bibitem[Allain(2020)]{allain2020}
R.~Allain.
\newblock Physics on ice.
\newblock \emph{Physics World}, 33\penalty0 (4):\penalty0 60, April 2020.

\bibitem[Brefeld et~al.(2019)Brefeld, Lasek, and Mair]{brefeld2019}
U.~Brefeld, J.~Lasek, and S.~Mair.
\newblock Probabilistic movement models and zones of control.
\newblock \emph{Machine Learning}, 108\penalty0 (1):\penalty0 127--147, 2019.

\bibitem[Fujimura and Sugihara(2005)]{fujimura2005geometric}
A.~Fujimura and K.~Sugihara.
\newblock Geometric analysis and quantitative evaluation of sport teamwork.
\newblock \emph{Systems and Computers in Japan}, 36\penalty0 (6):\penalty0
  49--58, 2005.

\bibitem[Lipps et~al.(2011)Lipps, Galecki, and Ashton-Miller]{lipps2011}
D.~B. Lipps, A.~T. Galecki, and J.~A. Ashton-Miller.
\newblock On the implications of a sex difference in the reaction times of
  sprinters at the beijing olympics.
\newblock \emph{PloS one}, 6\penalty0 (10):\penalty0 e26141, 2011.

\bibitem[Spearman(2016)]{spearman2016pitch}
W.~Spearman.
\newblock Quantifying pitch control, 02 2016.

\bibitem[Spearman et~al.(2017)Spearman, Basye, Dick, Hotovy, and
  Pop]{spearman2017physics}
W.~Spearman, A.~Basye, G.~Dick, R.~Hotovy, and P.~Pop.
\newblock Physics-based modeling of pass probabilities in soccer.
\newblock In \emph{Proceeding of the 11th MIT Sloan Sports Analytics
  Conference}, 2017.

\bibitem[Stokes(1850)]{stokes1850effect}
G.~Stokes.
\newblock On the effect of internal friction of fluids on the motion of
  pendulums.
\newblock \emph{Trans. Camb. phi1. S0c}, 9\penalty0 (8):\penalty0 106, 1850.

\bibitem[Taki and Hasegawa(2000)]{taki2000visualization}
T.~Taki and J.-i. Hasegawa.
\newblock Visualization of dominant region in team games and its application to
  teamwork analysis.
\newblock In \emph{Proceedings computer graphics international 2000}, pages
  227--235. IEEE, 2000.

\bibitem[{The Associated Press}(2019)]{cbcnews_2019}
{The Associated Press}.
\newblock {McDavid NHL's fastest skater but women's star Coyne Schofield steals
  show} | {CBC} sports, Jan. 26 2019.
\newblock URL
  \url{https://www.cbc.ca/sports/hockey/nhl/nhl-all-star-skills-competition-mcdavid-coyne-1.4993845}.

\bibitem[Voronoi(1908)]{voronoi1908nouvelles}
G.~Voronoi.
\newblock Nouvelles applications des param{\`e}tres continus {\`a} la
  th{\'e}orie des formes quadratiques. deuxi{\`e}me m{\'e}moire. recherches sur
  les parall{\'e}llo{\`e}dres primitifs.
\newblock \emph{Journal f{\"u}r die reine und angewandte Mathematik (Crelles
  Journal)}, 1908\penalty0 (134):\penalty0 198--287, 1908.

\bibitem[Williams and Rasmussen(2006)]{williams2006gaussian}
C.~K. Williams and C.~E. Rasmussen.
\newblock \emph{Gaussian processes for machine learning}, volume~2.
\newblock MIT press Cambridge, MA, 2006.

\end{thebibliography}
